\documentclass[aps,showpacs,twocolumn]{revtex4}
\usepackage{graphicx}
\begin{document}
\newcommand{\be}{\begin{equation}}
\newcommand{\en}{\end{equation}}
\title{A Dynamical Cosmological Term from the Verlinde's Maps}
\author{Luis Alejandro Correa-Borbonet}\email{borbonet@cpd.ufmt.br}
\affiliation{Departamento de F\'\i sica \\ Universidade Federal de Mato Grosso,\\
Av. Fernando Corr\^{e}a da Costa, $s/n^o$-Bairro Coxip\'{o}\\
78060-900-Cuiab\'{a}-MT, Brazil}

\begin{abstract}
In this letter it is proposed another generalization of the
Verlinde's maps for the case  $\Lambda
\neq 0$. Thermodynamical arguments combined with this proposal
conduce to a inverse square-law cosmological term behavior.
\end{abstract}
\pacs{98.80.-k, 11.25.Hf}

\maketitle
Since the original introduction of the cosmological cons\-tant
$\Lambda$ by Eins\-tein, ai\-ming to keep {\it his universe}
static, ``the genie has been let out of the bottle and it is no longer easy to force it back in",
paraphrasing Zel'dovich \cite{zeldovich}. The cosmological
constant has been an elusive pro\-blem during all these years of
research. Meanwhile, its interpretation as a measure of the energy
density of the vacuum has been an important issue that particle
theorists have realized.

Usually on textbooks and also in some subareas within Cosmology
research $\Lambda$ is considered as a cons\-tant parame\-ter.
However, recent de\-ve\-lop\-ments in particle physics and
inflationary theory have indicated that the cosmological term
should be treated as a dynamical quantity\cite{overduin}.

On the other hand during the recent years many stu\-dies have been
done about the Verlinde's maps \cite{Verlinde}. These very
interesting maps establish a relation between the Friedmann
Robertson Walker (FRW) equations that control the cosmological
expansion and the formulas that relate the energy and the entropy
of a Conformal Field Theo\-ry (CFT). In particular three amazing
relations mapping the $D$-dimensional Friedmann equation \be
H^{2}=\frac{16\pi G}{(D-1)(D-2)}\frac{E}{V}-\frac{1}{R^{2}}
\label{friedmann} \en into the Cardy's entropy formula
\cite{Cardy} \be S=2\pi
\sqrt{\frac{c}{6}(L_{0}-\frac{c}{24})},\label{eqn:cardy} \en well
known in two dimensional  CFT, have been set up. Those are
\begin{eqnarray}
2\pi L_{0} & \Rightarrow & \frac{2\pi}{D-1}ER , \nonumber \\
2\pi \frac{c}{12} & \Rightarrow & (D-2)\frac{V}{4GR}, \\
S  & \Rightarrow & (D-2)\frac{HV}{4G}, \nonumber
\label{verlindesmap}
\end{eqnarray}
where $L_0$ is the zero-mode Virasoro operator, $c$ the central
charge, $R$ the scale factor and $H=\dot{R}/R$ the Hubble
parameter with the dot representing the time derivative.

 The scenario considered in \cite{Verlinde}  was that of a closed
radia\-tion dominated FRW universe with a vanishing cosmological
constant. Within this context Verlinde proposed that the Cardy
formula for $2D$ CFT can be gene\-ralized to arbitrary spacetime
dimensions. Such a gene\-ralized entropy formula is known as the
Cardy-Verlinde formula.

Later soon, this result was studied and understood in several set
ups \cite{ivo,abda,wang,veneziano}.  Specifically in the ref.
\cite{abda} the maps were generalized to two different classes of
universes including cosmological constant: the de Sitter (dS)
closed and the Anti-de Sitter (AdS) flat. Both occupied by a
universe-sized black hole.  In the same way, working in the branes
context, the CFT dominated universe has been described as a
co-dimension one brane in the background of various kinds of (A)dS
black holes. In such cases, when the brane crosses the black hole
horizon, the entropy and temperature are expressed in terms of the
Hubble constant and its time derivative.  Such relations hold
precisely when the holographic entropy bound is saturated.

Verlinde's proposal has inspired a considerable activi\-ty
shedding further light on the various aspects of the
Cardy-Verlinde formula. Nevertheless there is still no answer to
the question about whether the merging of the CFT and FRW
equations is a  mere formal coincidence or, quoting Verlinde,
whether this fact ``{\it strongly indicates that both sets of
these equations arise from a single underlying fundamental
theory}''.

Returning to the Verlinde's maps (\ref{verlindesmap}), we would
like to stress that they are valid, without restrictions, for all
times, at least formally. Nevertheless up to now the study of
these maps has been restricted to the use of the machinery of
entropy bounds, black holes, branes and the holographic principle
\cite{susskind}. Therefore our aim is to follow the spirit of the
Verlinde's concern about the amazing relation between these two
equations and what is behind. This direction was followed in
\cite{tu}, where some mathematical properties of the Friedmann's
equation (\ref{friedmann}) that justified, at least in principle,
the relation with the Cardy's formula (\ref{eqn:cardy}) were
explored. In this letter we will show how it is possible to obtain
a generalization of the Verlinde's maps for the case of the
cosmological term di\-fferent from zero just expressing
$\Lambda$ in terms of the vacuum energy density and this will lead us to
find a dynamical behavior of the cosmological term.

In the presence of the cosmological term the 
Friedmann  equation takes the form,
 \be H^{2}=\frac{16\pi
G}{(D-1)(D-2)}\frac{E}{V}-\frac{1}{R^{2}}+\frac{\Lambda}{D-1}.
\label{friedmann22} \en
Now, in order to obtain the generalization of the Verlinde's maps we will just deal 
with the first relationship, i.e,
\be
2\pi L_{0}  \Rightarrow  \frac{2\pi}{D-1}ER . \nonumber \\
\en Our proposal is to shift the value of the energy $E$ adding
the vacuum energy $E_{vac}$. In this manner, the energy
$\mathcal{E}$ will be expressed as
 \be
\mathcal{E}=E+E_{vac},\label{eqn:newenergy}
 \en
 where
 \be
E_{vac}=\rho_{vac}V=\frac{(D-2)\Lambda}{16\pi G}V, \label{evac}
\en with $\rho_{vac}$ representing the vacuum energy density.
Therefore while the two last conditions of the Verlinde's maps
(\ref{verlindesmap}) are preserved, the Virasoro operator $L_0$ it
is redefined now in terms of ${\cal E}$. Thus, the Verlinde's maps
will take the form
\begin{eqnarray}
2\pi L_{0} & \Rightarrow & \frac{2\pi}{D-1}\mathcal{E}R , \nonumber \\
2\pi \frac{c}{12} & \Rightarrow & (D-2)\frac{V}{4GR}, \label{eqn:verlin2}\\
S  & \Rightarrow & (D-2)\frac{HV}{4G}, \quad \nonumber
\end{eqnarray}
and can be easily checked that the $D$-dimensional Friedmann
(\ref{friedmann22}) equation turns into the Cardy formula.

At this point it is important to stress that we conside\-red the
maps (\ref{eqn:verlin2}) valid for all times. Particularly we will
think $c$ as a $c$-function. This choice is based in some
arguments given by Strominger in \cite{strom} where was conjectured
that the cosmological evolution of an $n+1$ dimensional universe
has a dual representation as the renormalization group(RG) flow
between two conformal fixed points of a $n$-dimensional Euclidean
field theory. The RG flow begins at a UV(ultraviolet) conformally
invariant fixed point and ends at an IR(infrared) conformally
invariant fixed point. Since late(early) times corresponds to the
UV(IR) fixed point then the RG flow corresponds to evolution back
in time from the future to the past \cite{strom}. The proposed
$c$-function was \be c \sim
\frac{1}{G_{N}|\frac{\dot{a}}{a}|^{n-1}}.\label{eqn:function} \en
Using the Einstein equations can be showed that
$\partial_{t}(\dot{a}/a)<0$, provided that any matter in the
spacetime satisfies the null energy condition. In other words,
this guarantees that the c-function will always decrease in a
contracting phase of the evolution or increase in an expanding
phase\cite{myers}. Also, if the $c$-function can be eva\-luated on
each slice of some foliation of the spacetime and the slice can be
embedded in some de Sitter space the $c$-function
(\ref{eqn:function}) takes the form \be c \sim
\frac{1}{G_{N}\Lambda^{(n-1)/2}} \label{eqn:clambda} \en In the
originals papers were considered spatially flat cosmological models
with $k=0$ but their $c$-function was subsequently generalized
\cite{myers}, \cite{thorlacius} so as to apply to $k \neq 0 $
models as well.

Returning to our main line we remember that in \cite{Verlinde} was
also obtained a universal Cardy entropy formula in terms of the
energy $E$ and the Casimir energy $E_C$. It should be also noted
that in a CFT with large central charge the entropy and energy are
not purely extensive. Being the volume finite, the energy of a CFT
contains a non-extensive Casimir contribution proportional to the
central charge $c$. Therefore, taking into account this fact, the
total energy can be decomposed as, \be
 E(S,V)=E_{E}(S,V)+\frac{1}{2}E_{C}(S,V), \label{eqn:descom}
\en
 where the first term is related to the purely extensive part
of the energy and the second term represents the Casimir energy.

On the other hand it is known that conformal inva\-riance implies
that the product $ER$ is independent of the volume and it is only
a function of the entropy $S$. Considering this fact and the
behavior of the extensive and sub-extensive parts of the energy
can be found that
\begin{eqnarray}
E_{E}=\frac{a}{4\pi R} S^{1+1/(D-1)},
\;\;\;\;\;\;\;\;\;E_{C}=\frac{b}{2\pi R} S^{1-1/(D-1)},
\end{eqnarray}
where $a$ and $b$ are a priori arbitrary positive coefficients.
Using these expressions the entropy $S$ takes the form
\be
S=\frac{2\pi R}{\sqrt{ab}}\sqrt{E_{C}(2E-E_{C})}.
 \en
Obviously the formula (\ref{eqn:cardy}) can be recovered, ignoring
the normalization, if we insert $ER=L_{0}$ and $E_{C}R=c/12$.

Extending the above mentioned reasoning to the case $\Lambda> 0$
it can be realized that it is much more natural to apply here the
decomposition of the energy (\ref{eqn:descom}) since the Casimir
energy is often invoked, via vacuum energy, as a decisive factor
for explaining the cosmological term. So, \be
\mathcal{E}=E+\frac{1}{2}E_{C}. \en Therefore we arrive at the
following relations
\begin{eqnarray}
E_{E}=E, \;\;\;\;\;\;\;\;\; E_{C}=2E_{vac}.
\end{eqnarray}
Writing the last expression explicitly we get, \be
\frac{c}{12R}=\frac{2(D-2)\Lambda}{16\pi G}V.
 \en

Next we found, with the help of the second relation of the
Verlinde's maps, that the cosmological term behaves as
\be \Lambda=\frac{1}{R^{2}}\label{eqn:cosmoterm} \, ,
 \en
a decay law for $\Lambda$!(or vacuum energy density decay!) That is a remarkable result since among
the decay laws of the cosmological term in function of the scale
factor ($ \Lambda=\alpha R^{-m}$ with $\alpha$ being a constant
parameter), proposed in the literature, this case ($m=2$) has been
the one that has received most of the attention\cite{overduin}.
The inverse-square law dependence is supported by dimensional
\cite{chen} and phenomenological arguments \cite{vish}. In our
scenario this behavior has appeared 1-) as a consequence of
identifying the vacuum energy with half the Casimir energy and 2-)
as a consequence of the use of the relations between the
cosmological quantities and conformal quantities through the
Verlinde's maps. Here it is worth to note that if we introduce
(\ref{eqn:cosmoterm}) in the second map of (\ref{eqn:verlin2}) the
$c$-function (\ref{eqn:clambda}) is obtained. In other words, the
 behavior found for the cosmological term is compatible with the $c$-function proposed in \cite{myers}, \cite{thorlacius}.

On the other hand on the basics of a rich body of astronomical
observations\cite{observations,observations2} there is now convincing evidence
that the recent Universe is dominated by an exo\-tic dark energy
density with negative pressure, responsible for the cosmic acceleration. The simplest candidate for dark
energy is the cosmological constant. But, if ge\-neral relativity is correct, cosmic acceleration implies that there
must be a dark energy density which dimi\-nishes relatively slowly as the universe expands\cite{carroll}. In this context it has been studied the possibility that the dark
energy may decay\cite{carroll, varun, starobinsky} . Therefore the result obtained is also in agreement with the expectations from the dark energy side.

Thus, here we have gone one step ahead. Although we still do not
know how to calculate $\Lambda$ from first principles in this
approach we have generated a genuinely $R^{-2}$ varying $\Lambda$
law.

\begin{acknowledgments}
I would like to thank the Perimeter Institute as well as the
University of Waterloo for the kind hospitality where the last
version was done. I thank Achim Kempf, Robert Myers, Justin Khoury
for the comments and suggestions, and specially Freddy Cachazo for
also indicating the reference \cite{strom}. Also I acknowledge
 I. Cabrera-Carnero for useful discussions and suggestions and Sandro S. e Costa for indicating
 some references. This work has been
 supported by CNPq-FAPEMAT/UFMT. I wish to thank the anonymous referee for useful comments which helped to improve this work.
\end{acknowledgments}

\end{document}